\documentclass[runningheads]{llncs}
\usepackage{graphicx}
\usepackage{amsmath}
\usepackage{amssymb}
\DeclareMathOperator*{\argmin}{arg\,min}
\DeclareSymbolFont{matha}{OML}{txmi}{m}{it}
\DeclareMathSymbol{\varv}{\mathord}{matha}{118}
\usepackage{array}
\usepackage{tabularx}
\usepackage{multirow, booktabs}
\usepackage{float}
\usepackage{bm}
\usepackage[breaklinks=true,bookmarks=false]{hyperref}

\begin{document}
%
\title{Unsupervised Deformable Image Registration with Absent Correspondences in Pre-operative and Post-Recurrence Brain Tumor MRI Scans}
%
\titlerunning{Unsupervised Deformable Image Registration with Absent Correspondences}
%
%
%

\author{Tony C. W. Mok \and Albert C. S. Chung}

\institute{
	Department of Computer Science and Engineering,\\
	The Hong Kong University of Science and Technology, Hong Kong\\
	\email{\{cwmokab,achung\}@cse.ust.hk}
}

\maketitle              
\begin{abstract}
Registration of pre-operative and post-recurrence brain images is often needed to evaluate the effectiveness of brain gliomas treatment. While recent deep learning-based deformable registration methods have achieved remarkable success with healthy brain images, most of them would be unable to accurately align images with pathologies due to the absent correspondences in the reference image. In this paper, we propose a deep learning-based deformable registration method that jointly estimates regions with absent correspondence and bidirectional deformation fields. A forward-backward consistency constraint is used to aid in the localization of the resection and recurrence region from voxels with absence correspondences in the two images. Results on 3D clinical data from the BraTS-Reg challenge demonstrate our method can improve image alignment compared to traditional and deep learning-based registration approaches with or without cost function masking strategy. The source code is available at \url{https://github.com/cwmok/DIRAC}. 

\end{abstract}

\section{Introduction}
Registration of pre-operative and post-recurrence brain MRI images plays a significant role in discovering accurate imaging markers and elucidating imaging signatures for aggressively infiltrated tissue, which are crucial to the treatment plan and diagnosis of intracranial tumors, especially brain gliomas \cite{heiss2011multimodality,price2007predicting}. To better understand the location and extent of the tumor and its biological activity after resection, pre-operative and follow-up structural brain MRI scans of a patient first need to be aligned accurately. However, deformable registration between the pre-operative and follow-up scans, including post-resection and post-recurrence, is challenging due to possible large deformations and absent correspondences caused by tumor's mass effects \cite{dean1990gliomas}, resection cavities, tumor recurrence and tissue relaxation in the follow-up scans. 

Conventional registration methods mostly deal with the absent correspondence issue by (1) excluding the similarity measure of pathological regions \cite{brett2001spatial,clatz2005robust}, 2) replacing the pathological images with quasi-normal appearance \cite{yang2016registration,han2017efficient,kwon2015estimating} or 3) joint registration and segmentation framework \cite{risholm2009non,chitphakdithai2010non}. Excluding the pathological regions often requires manual delineation \cite{brett2001spatial} or initial seed \cite{chitphakdithai2010non,kwon2013portr} of the tumor regions in brain scans, which are often prohibitive and daunting to acquire in terms of labour cost and resources. Replacing the pathological image with the quasi-normal appearance, alternately, avoids the prerequisite of a prior pathological segmentation. However, modeling the tumor-to-quasi-normal appearance with a statistical model \cite{han2017efficient,kwon2015estimating} often requires extra image scans, i.e., image scans from a healthy population. Moreover, existing approaches based on quasi-normal images require accurate registration to a common atlas space for quasi-normal reconstruction. Ironically, accurate alignment with images suffered from mass effect is very hard to achieve without reconstruction. Therefore, the registration and reconstruction problems with quasi-normal approaches need to be interleaved in a costly iterative optimization process. Alternatively, an unsupervised approach \cite{risholm2009non} to accommodate resection and retraction of tissue was proposed for registering pre-operative and intra-operative brain images. Their method alternates between registering the brain scans using the demons algorithm with an anisotropic diffusion smoother and segmenting the resection using level set method in the space with high image intensity disagreement. Chitphakdithai \textit{et al.} \cite{chitphakdithai2010non} extended this idea to a simultaneous registration and resection estimation approach with the expectation-maximization algorithm and a prior on post-resection image intensities. Nevertheless, these methods rely on the costly iterative optimization, which can be up to $\sim3.5$ hours per case \cite{kwon2013portr}.

While recent deep learning-based deformable registration (DLDR) methods have achieved remarkable registration speed and superior registration accuracy \cite{balakrishnan2018unsupervised,dalca2018unsupervised,kim2019unsupervised,hu2019dual,heinrich2019closing,mok2020fast,mok2020large,mok2021conditional}, these registration algorithms are incapable of accurately registering pre-operative and post-recurrence images due to the absent correspondence problem. A learning-based registration method for images with pathology was presented in \cite{han2020deep} which dealt with missing correspondence by joint estimating the vector-momentum parameterized stationary velocity field (vSVF) and quasi-normal image to drive the registration. Nevertheless, the reconstruction of the quasi-normal image requires explicit tumor segmentation in the training phase. Moreover, the large deformation caused by the mass effect of tumor is difficult to model without resorting to complex multi-stage warping pipelines.

In this paper, we present an unsupervised joint registration and segmentation learning framework, in which a large deformation image registration network and a forward-backward consistency constraint are leveraged to estimate the valid and absent correspondence regions along with the dense deformation fields in a bidirectional manner, for pre-operative and post-recurrence registration. Instead of using a manual delineation or image intensity disagreement to segment the pathological regions, our method leverages the forward-backward consistency constraint of the bidirectional deformation fields to explicitly locate regions with absent correspondence and excludes them in the similarity measure in an unsupervised manner. We present extensive experiments with a pre-operative and post-recurrence brain MR dataset, demonstrating that our method achieves accurate registration accuracy in brain MR scans with pathology.

\section{Methods}
\begin{figure}[t]
	\centering
	\includegraphics[width=1.0\linewidth]{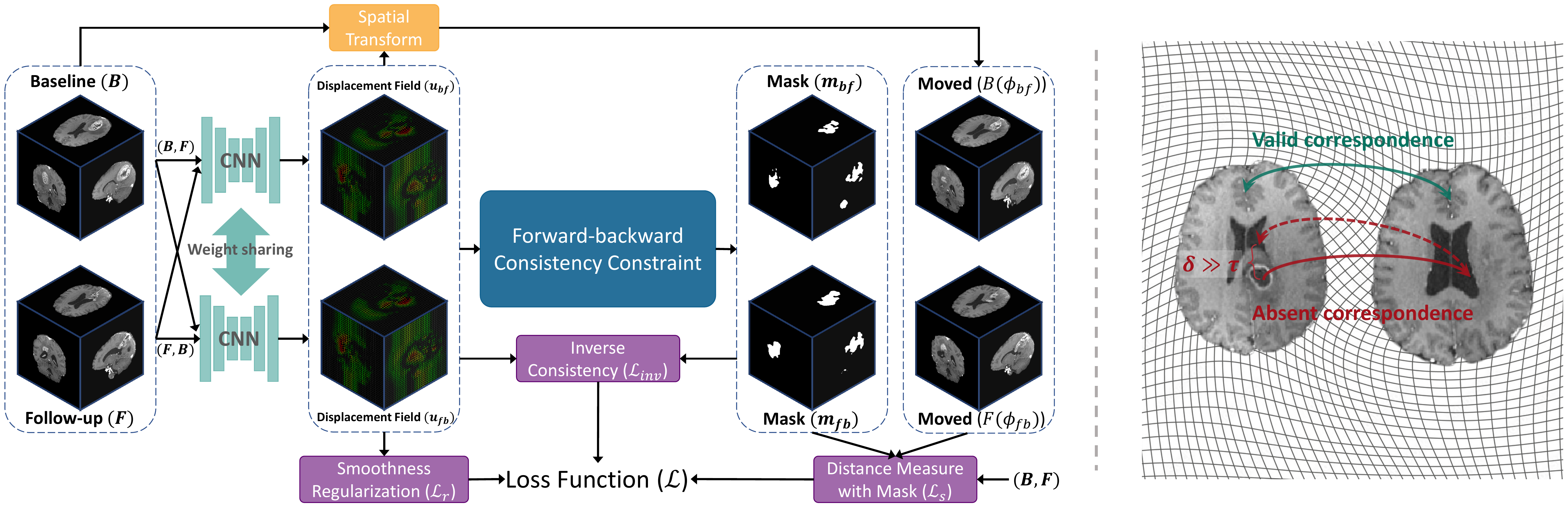}
	\caption{Overview of the proposed method (Left) and the semantic representation of the forward-backward consistency constraint (Right). Our method jointly estimates the bidirectional deformation fields and locates regions with absent correspondence (denoted as mask). The regions with absent correspondence are excluded in the similarity measure during training. For brevity, the magnitude loss of the masks is omitted in the figure.} \label{fig:FCN_archi}
\end{figure}

Our goal is to establish a dense non-linear correspondence between the pre-operative scan and the post-recurrence scan of the same subject, where regions without valid correspondence are excluded in the similarity measure during optimization. Our method builds on the previous DLDR method \cite{mok2021conditional} and extends it to accommodate the absent correspondence issue in the pre-operative and post-recurrence scans. 

\vspace{-10pt}
\subsection{Bidirectional Deformable Image Registration}
Let $B$ and $F$ be the pre-operative (baseline) scan $B$ and post-recurrence (follow-up) scan defined over a $n$-D mutual spatial domain $\Omega \subseteq \mathbb{R}^n$. In this paper, we focus on 3D deformable registration, i.e., $n = 3$ and $\Omega \subseteq \mathbb{R}^3$ and assume that $B$ and $F$ are affinely aligned to a common space.

Figure \ref{fig:FCN_archi} depicts an overview of our method. We parametrize the deformable registration problem as a bidirectional registration problem $\bm{u}_{bf} = f_\theta(B, F)$ and $\bm{u}_{fb} = f_\theta(F, B)$ with CNN, where $\theta$ is a set of learning parameters and $\bm{u}_{bf}$ represents the displacement field that transform $B$ to align with $F$, i.e., $B(x+\bm{u}_{bf}(x))$ and $F(x)$ define similar anatomical locations for each voxel $x\in\Omega$ (except voxels with absent correspondence). The proposed method works with any CNN-based DLDR methods. In order to accommodate the large deformation and variation of anatomical structures caused by the tumor’s mass effect, we parametrize an example of the function $f_\theta$ with the conditional deep Laplacian pyramid image registration network (cLapIRN) \cite{mok2021conditional}, which is capable of large deformation and rapid hyperparameter tuning for the smoothness regularization in a wide range of applications \cite{hering2021learn2reg}. Despite the multi-resolution optimization strategy used in the cLapIRN, vanilla cLapIRN is incapable of accurately registering images with absent correspondence, i.e., missing correspondence caused by the tumor resection and recurrence, edema and cavity. Therefore, instead of measuring the similarity of $B$ and $F$ for every voxel $x \in \Omega$, our method estimates the regions with absent correspondence in both $B$ and $F$ domains using the bidirectional displacement fields and the forward-backward consistency constraint, and only measures the similarity on regions with valid correspondence during optimization.

\subsection{Forward-Backward Consistency Constraint}
Conventionally, regions with absent correspondence can be detected by comparing the appearance or image intensities of the warped scan to the target scan or an atlas \cite{risholm2009non,kwon2015estimating}. However, corresponding regions in the pre-operative and post-recurrence scans may have different intensity profiles, which make their approaches less robust in practice. Therefore, we depart from approaches with spatial prior and extend the forward-backward consistency \cite{sundaram2010dense,wang2018occlusion,liu2019selflow,meister2018unflow} instead. We design a forward-backward consistency constraint to locate regions with absent correspondence in the baseline and follow-up scans. The forward-backward (inverse consistency) error $\delta_{bf}$ from $B$ to $F$ is defined as:

\begin{equation}\label{eq:forward_backward}
	\delta_{bf}(x) = |\bm{u}_{bf}(x) + \bm{u}_{fb}(x+\bm{u}_{bf}(x))|_2.
\end{equation}

We estimate the regions with absent correspondence by checking the consistency of the forward and backward displacement fields. For any voxel $x$, if there is a significant violation of inverse consistency in x, i.e., $\delta_{bf}(x)>\tau_{bf}$, the voxel x is either without valid correspondence or the displacement field is not accurately estimated. $\tau_{bf}$ is the pre-defined threshold and is defined as follows:

\begin{equation}\label{eq:threshold}
	\tau_{bf} = \sum_{x\in\{x | F(x)>0\}} \frac{1}{N_f} \big(|\bm{u}_{bf}(x) + \bm{u}_{fb}(x+\bm{u}_{bf}(x))|_2\big) + \alpha,
\end{equation}

\noindent where the first term grants a tolerance interval that allows estimation errors to increase with the overall complexity of the registration and $\alpha$ is a constant. Then, we create a binary mask $\bm{m}_{bf}$ to mark voxels with absent correspondence as follows: 

\begin{equation}\label{eq:binary_mask}
    \bm{m}_{bf}(x)= 
\begin{cases}
    1,& \text{if } (\bm{A} \star \delta_{bf})(x) \geq \tau_{bf}\\
    0,              & \text{otherwise}
\end{cases}
\end{equation}

\noindent where $\bm{A}$ denotes an averaging filter of size $(2p+1)^3$ and $\star$ denotes a convolution operator with zero-padding $p$. Since the estimated registration fields will fluctuate during learning, we apply an averaging filter to the estimated forward-backward error to stabilize the estimation of the binary mask as well as to alleviate the effect of outliners to the mask estimation. For the mask $\bm{m}_{fb}$ in the backward to forward direction, we can define it in a symmetric way with $\bm{u}_{fb}$ and $\bm{u}_{bf}$ exchanged. We set $\alpha=0.015$ and $p=4$ in all our experiments. The values of $\alpha$ and $p$ are determined by measuring the forward-backward error of the pathological regions from a vanilla cLapIRN model.

\subsection{Inverse Consistency}
Since the decision of regions with absent correspondence is highly dependent on the inverse consistency error in our method, we further enforce the inverse consistency on the regions with valid correspondence. Mathematically, the inverse consistency loss $\mathcal{L}_{\text{inv}}$ is defined as:

\begin{equation}\label{eq:inv_loss}
    \mathcal{L}_{\text{inv}}=\sum_{x\in\Omega} (\delta_{bf}(x)(1-\bm{m}_{bf}(x)) + \delta_{fb}(x)(1-\bm{m}_{fb}(x))),
\end{equation}
\noindent where the measure of inverse consistency error $\delta$ is masked with the regions with valid correspondence $(1-\bm{m})$ via elementwise multiplication.

\begin{figure}[t]
	\begin{center}
		\includegraphics[width=1.0\linewidth]{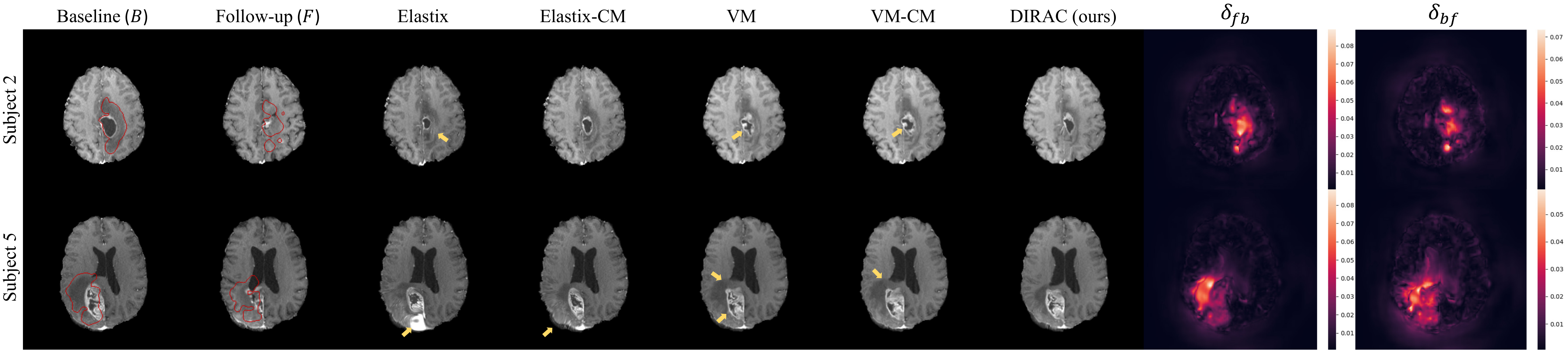}
	\end{center}
	
	\caption{Example axial T1ce MR slices of resulting warped images ($B$ to $F$) from the baseline methods and our proposed method. Registration artefacts are highlighted with yellow arrows. The forward-backward errors ($\delta_{fb}$ and $\delta_{bf}$) of our method are shown next to our result. The estimated regions with absent correspondence from our method are overlaid with the baseline and follow-up scans (in red).}\label{fig:qualitative}

\end{figure}

\subsection{Objective Function}
Given the deformation fields $\phi_{bf} = Id + \bm{u}_{bf}$ and $\phi_{fb} = Id + \bm{u}_{fb}$, where $Id$ is the identity transform. The objective of our proposed method is to compute the optimal deformation fields that minimize the dissimilarity measure of $B(\phi_{bf})$ and $F$ as well as $B$ and $F(\phi_{fb})$ in regions with valid correspondence. Specifically, we adopt the negative local cross-correlation (NCC) with masks to exclude the similarity measure of regions without valid correspondence as shown in eq. \ref{eq:sim_loss}.

\begin{equation}\label{eq:sim_loss}
    \mathcal{L}_{\text{s}}=-\text{NCC}(F, B(\phi_{bf}), (1-\bm{m}_{bf})) -\text{NCC}(B, F(\phi_{fb}), (1-\bm{m}_{fb})).
\end{equation}

To encourage smooth solution and penalize implausible solutions, we adopt a diffusion regularizer:
\begin{equation}\label{eq:reg_loss}
    \mathcal{L}_{\text{r}}=||\nabla \bm{u}_{bf}||^2_2+||\nabla \bm{u}_{fb}||^2_2.
\end{equation}

\noindent Hence, the complete loss function is therefore:
\begin{equation}\label{eq:objective}
	\mathcal{L} = (1-\lambda_{reg})\mathcal{L}_{\text{s}} + \lambda_{reg}\mathcal{L}_{\text{r}} + \lambda_{inv}\mathcal{L}_{\text{inv}} + \frac{\lambda_{m}}{N}(|\bm{m}_{bf}|_1+|\bm{m}_{fb}|_1),
\end{equation}

\noindent where $\lambda_{reg}$, $\lambda_{inv}$ and $\lambda_{m}$ are the hyperparameters to balance the loss functions. $N$ denotes the number of voxels in the mutual spatial domain $\Omega$ and the last term is to avoid the trivial solution where all voxels are marked in $\bm{m}_{bf}$ and $\bm{m}_{fb}$. During training, we follow the conditional registration framework in \cite{mok2021conditional} to sample $\lambda_{reg} \in [0,1]$ and set $\lambda_{reg} = 0.3$ in the inference phase. Formally, the optimal learning parameters $\theta^*$ is estimated by minimizing the complete loss $\mathcal{L}$ function using a training dataset $D$, as follows:

\begin{equation}\label{eq:training}
\begin{split}
	\theta^* =& \argmin_\theta \Big[ \mathbb{E}_{(B,F) \in D} \; \mathcal{L}\big(B, F, \bm{u}_{bf}, \bm{u}_{fb}, \bm{m}_{bf}, \bm{m}_{fb}\big) \Big].
\end{split}
\end{equation}

\vspace{-12pt}
\section{Experiments}

\begin{figure}[t]
	\centering
	\begin{tabular}{cc}
		\includegraphics[width=0.5\linewidth]{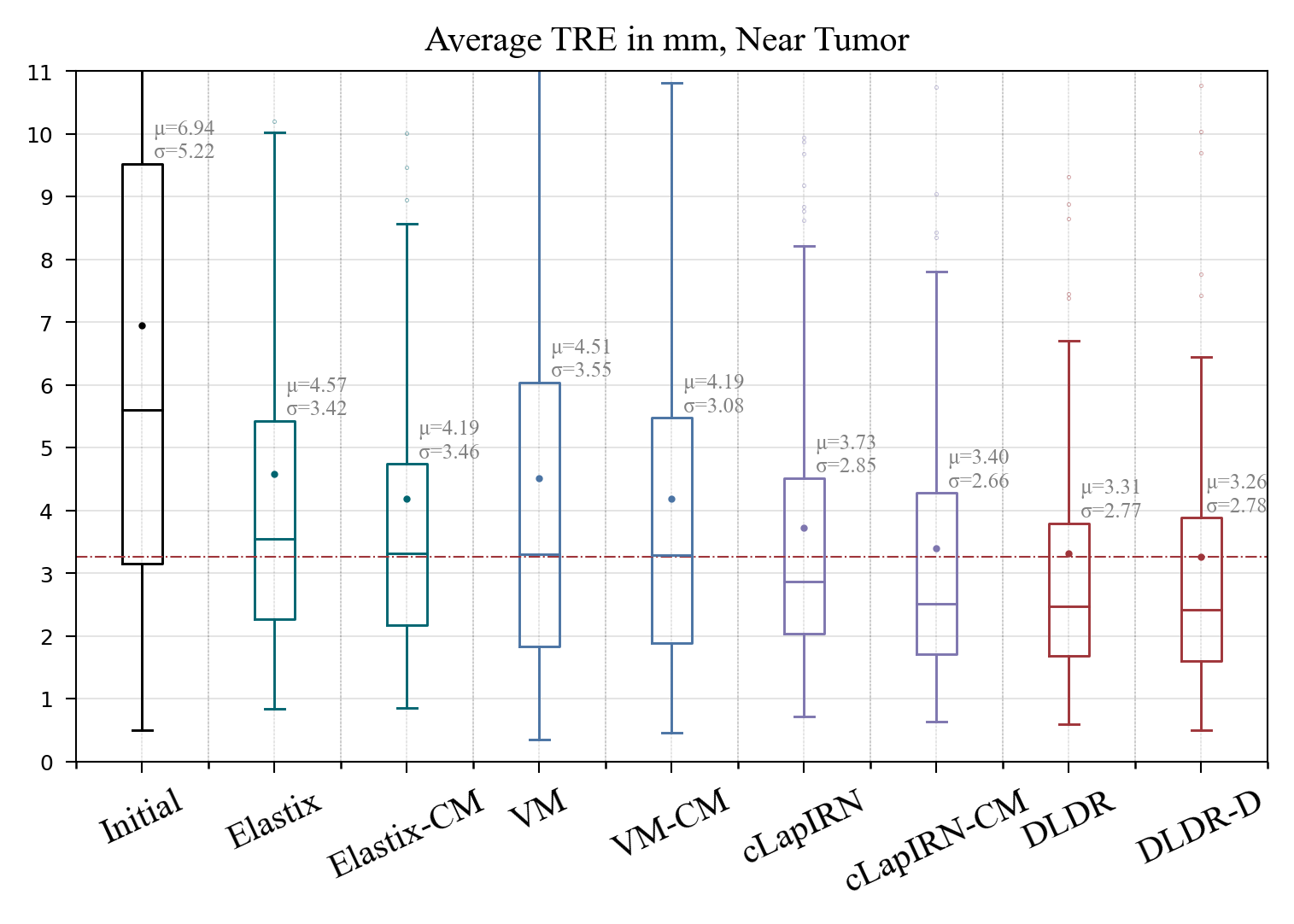} &
		\includegraphics[width=0.5\linewidth]{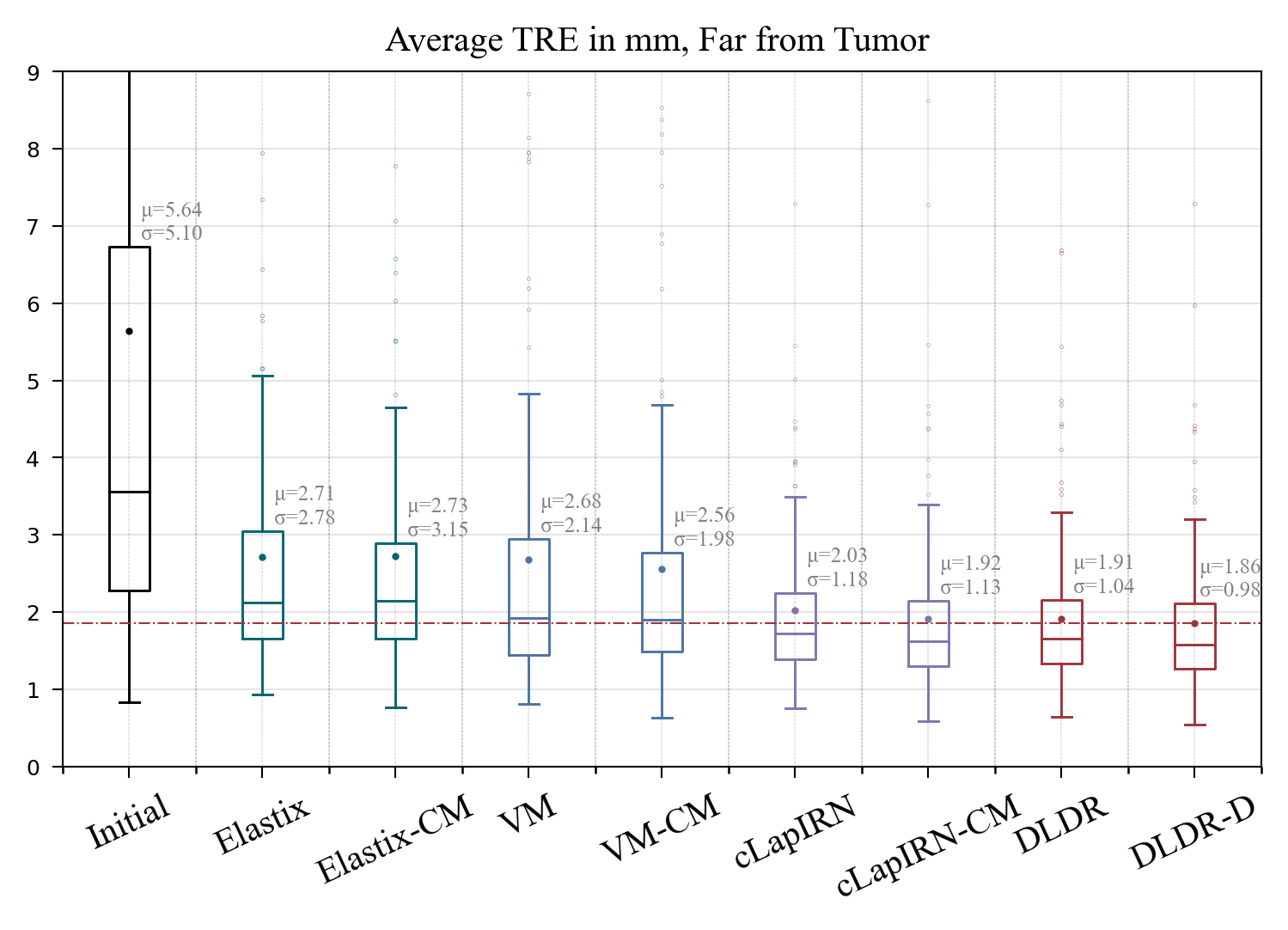} \\
	\end{tabular}
	
	\caption{Boxplots illustrate that the average target registration error (TRE) near tumor (left) and far from tumor (right). The mean ($\mu$) and standard deviation ($\sigma$) are shown next to the 75$^{th}$ percentile of each box.} \label{fig:results}
	\vspace{-12pt}

\end{figure}

\subsubsection{Data and Pre-processing}
We evaluate our method on the brain tumor MR registration task using the 3D clinical dataset from the BraTS-Reg challenge \cite{baheti2021brain}, which consists of 160 pairs of pre-operative and follow-up brain MR scans of glioma patients taken from different timepoint. Each timepoint contains native T1, contrast-enhanced T1-weighted (T1ce), T2-weighted and FLAIR MRI. 140 pairs of scans are associated with 6 to 50 manual landmarks in both scans and 20 scans with landmarks in the follow-up scan only. All scans have carried out standard processing, including skull stripping, affine spatial normalization and resampled to the $1 \ \text{mm}^3$ isotropic resolution. We use the DeepMedic \cite{kamnitsas2017efficient} to segment the tumor core in each pre-operative scan. The tumor segmentation map is used in cost function masking for baseline methods. For learning-based methods, we further resample the scans to size of $160\times160\times80$ with $1.5\times1.5\times1.94 \ \text{mm}^3$ isotropic resolution in the training phase and upsample the solutions to $1 \ \text{mm}^3$ isotropic resolution with bilinear interpolation in the evaluation. We perform 5-fold cross-validation and divide the 140 pairs of scans into 5 folds with equal size. In each group, we join 4 folds of data and the additional 20 pairs of scans as training set and validation set, and 1 fold as the test set. Specifically, for each group, we split the dataset into 122, 10, and 28 cases for training, validation and test sets.

\vspace{-12pt}
\subsubsection{Implementation}
Our proposed method and the other baseline methods are implemented with PyTorch 1.9 and deployed on the same machine, equipped with an Nvidia Titan RTX GPU and an Intel Core (i7-4790) CPU. We build our method on top of the official implementation of 3-level cLapIRN with default parameters available in \cite{offical_clapirn}. We set $\lambda_{reg}$, $\lambda_{inv}$ and $\lambda_{m}$ to 0.3, 0.5 and 0.01, respectively. We use Adam optimizer with a fixed learning rate 0.0001. All learning-based methods are trained from scratch. 

\vspace{-12pt}
\subsubsection{Measurement}
We register each pre-operative scan to the corresponding follow-up scan of the same patient, propagate the landmarks of the follow-up scan using the resulting deformation field and measure the mean target registration error (TRE) of the paired landmarks with Euclidean distance in millimetres. We divide the landmarks into two sets: 1) landmarks within 30mm from the tumor region (Near tumor), and 2) landmarks outside the 30mm tumor region (Far from tumor), using tumor segmentation maps and morphological dilation. We further measure the robustness of the registration. We follow \cite{baheti2021brain} to define the robustness for a pair of scans as the relative number of successfully registered landmarks. As the local deformation at voxel $p$ is invertible if and only if the Jacobian determinant of $p$ ($|J_\phi|(p)$) is larger than zero, we also measure the number of percentage of the voxels with Jacobian determinant smaller or equal to 0 (denoted as $\%|J_\phi|_{\leq0}$). We also measure the elapsed time in seconds for computations of each case in the inference phase ($\text{T}_\text{test}$).

\vspace{-12pt}

\subsubsection{Baseline Methods}
We compare our method (denotes as DIRAC) with a conventional approach (denoted as Elastix\cite{klein2009elastix}) and two cutting edge DLDR methods (denoted as VM\cite{balakrishnan2018unsupervised} and cLapIRN\cite{mok2021conditional}). For Elastix, we use the official implementation in the SimpleElastix library \cite{marstal2016simpleelastix}, which includes a 3-level iterative optimization scheme. For VM and cLapIRN, we use their official implementations with the best parameters reported in their papers. We also report the results of methods with cost function masking using the tumor core segmentation map for each method (denoted with postfix -CM). Note that the cost function masking strategy in learning-based methods is defined as excluding the similarity measure of the tumor region during the training phase, and the tumor segmentation is hidden during the inference phase, as opposed to conventional methods. All DLDR methods are trained from scratch with T1ce MR scans as input, except for our variant (denoted as DIRAC-D), which employs both the T1ce and T2-weighted scans of each case as input. 

\vspace{-12pt}
\subsubsection{Results and Discussions}
\begin{table}[t]
	\centering
	\caption{Quantitative results of the pre-operative and follow-up brain MR registration. Results are provided as mean$\pm$(standard deviation) Initial: spatial normalization. Runtime result highlighted with a asterisk ($^*$) denotes runtime with CPU only. To our knowledge, Elastix does not have a GPU implementation. $\uparrow$: higher is better, and $\downarrow$: lower is better.}
	\label{tab:result}
	\resizebox{\textwidth}{!}{%
		\begin{tabular}{ccccccccc}
			\toprule[1.5pt]
			\multirow{2}{*}{Method} & \multicolumn{3}{c}{Near Tumor} & \multicolumn{3}{c}{Far from Tumor} & \\
			\cmidrule(lr){2-4}\cmidrule(lr){5-7}
			& \rule{1pt}{0ex} TRE$\downarrow$ & Robustness$\uparrow$ & $\%|J_\phi|_{\leq0}\downarrow$ & \rule{1pt}{0ex} TRE$\downarrow$ & Robustness$\uparrow$ & $\%|J_\phi|_{\leq0}\downarrow$ & $\text{T}_\text{test} (\text{sec})$$\downarrow$\\
			\midrule[1pt]
			Initial \hspace{0.1cm} & $6.94\pm(5.22)$ & $-$ & $-$ & $5.64\pm(5.10)$ & $-$ & $-$ & $-$ \\
			\midrule[1pt]
            Elastix-CM \hspace{0.1cm} & $4.19\pm(3.46)$ & $0.71\pm(0.34)$ & $0.16\pm(0.89)$ & $2.73\pm(3.15)$ & $0.72\pm(0.31)$ & $0.08\pm(0.48)$ & $120.17$*$\pm(6.31)$ \\
            VM-CM \hspace{0.1cm} & $4.19\pm(3.08)$ & $0.82\pm(0.27)$ & $0.32\pm(0.42)$ & $2.56\pm(1.98)$ & $0.82\pm(0.23)$ & $0.27\pm(1.53)$ & $0.069\pm(0.001)$ \\
            cLapIRN \hspace{0.1cm} & $3.73\pm(2.85)$ & $0.80\pm(0.30)$ & $0.51\pm(0.45)$ & $2.03\pm(1.18)$ & $0.80\pm(0.25)$ & $0.21\pm(0.10)$ & $0.023\pm(0.004)$ \\
            cLapIRN-CM \hspace{0.1cm} & $3.40\pm(2.66)$ & $0.82\pm(0.28)$ & $0.83\pm(0.67)$ & $1.92\pm(1.13)$ & $0.82\pm(0.24)$ & $0.32\pm(1.59)$ & $0.024\pm(0.005)$ \\
            \midrule[1pt]
            DIRAC \hspace{0.1cm} & $3.31\pm(2.77)$ & $0.82\pm(0.28)$ & $0.18\pm(0.20)$ & $1.91\pm(1.04)$ & $0.82\pm(0.24)$ & $0.12\pm(0.24)$ & $0.023\pm(0.005)$ \\
            DIRAC-D \hspace{0.1cm} & $\bm{3.26}\pm(2.78)$ & $\bm{0.83}\pm(0.27)$ & $\bm{0.13}\pm(0.16)$ & $\bm{1.86}\pm(0.98)$ & $0.82\pm(0.25)$ & $0.08\pm(0.15)$ & $0.025\pm(0.005)$ \\
            
			\bottomrule[1.5pt]
		\end{tabular}
	}
\end{table}


Fig. \ref{fig:results} illustrates the box-and-whisker plots of average TRE of registered landmarks based on landmarks inside the 30 mm tumor boundary (Group 1) in the left graph as well as the one for the remaining landmarks in the right graph across the 140 subjects (Group 2). Among deformable methods with single MR modality as input, our method DIRAC has the lowest mean registration error of 3.31 and 1.91 mm in groups 1 and 2, respectively, which improves the registration error of our baseline method cLapIRN significantly by 0.42 mm (-11\%) and 0.17 mm (-8\%) in groups 1 and 2, respectively. Among the alternative methods, methods with cost function masking (-CM) show significant improvement over their baseline method in group 1 and the improvement gain in group 2 is less significant, suggesting that implicitly or explicitly enforcing the smooth deformations inside the masked tumor regions is effective to the registration near the tumor regions. Table \ref{tab:result} shows a comprehensive summary of the registration error, robustness, local invertibility and runtime results across the 140 subjects. As opposed to the alternative methods using cost function masking, our methods (DIRAC and DIRAC-D) have achieved the best overall results in a fully unsupervised manner without sacrificing the runtime advantage of learning-based methods. Comparing the results of DIRAC and DIRAC-D, our variant DIRAC-D, which leverages additional MR modality, slightly improves the registration error by 1.5\% and 2.6\% in groups 1 and 2, respectively. Fig. \ref{fig:qualitative} shows qualitative examples of the registration results for each method and the estimated regions with absent correspondence by our method. The results demonstrate our method is capable of accurately locating the regions without valid correspondence, i.e., the tumor and cerebral edema in the baseline scan of subject 2, and explicitly excluding these regions in similarity measure during the training phase further reduces artefacts in the patient-specific registration. 

\vspace{-12pt}
\section{Conclusion}
We have proposed a unsupervised deformable registration method for the pre-operative and post-recurrence brain MR registration, which capable of joint registration and segmentation of regions with absent correspondence. We introduce a novel forward-backward consistency constraint and a pathological-aware symmetric loss function. Compared to existing deep learning-based methods, our method addresses the absent correspondence issue in patient-specific registration and shows significant improvement in registration accuracy near the tumor regions. Compared to conventional methods, our method inherits the runtime advantage from deep learning-based approaches and does not require any manual interaction or supervision, demonstrating immense potential in the fully-automated patient-specific registration.

\bibliographystyle{splncs04}
\bibliography{myref}

\clearpage
\begin{appendix}
\renewcommand{\thesection}{\Alph{section}.}%

\vspace{-10pt}
\section{Example MR slices in axial, sagittal and coronal plane}
\vspace{-15pt}
\begin{figure}[h]
	\begin{center}
		\includegraphics[width=1.0\linewidth]{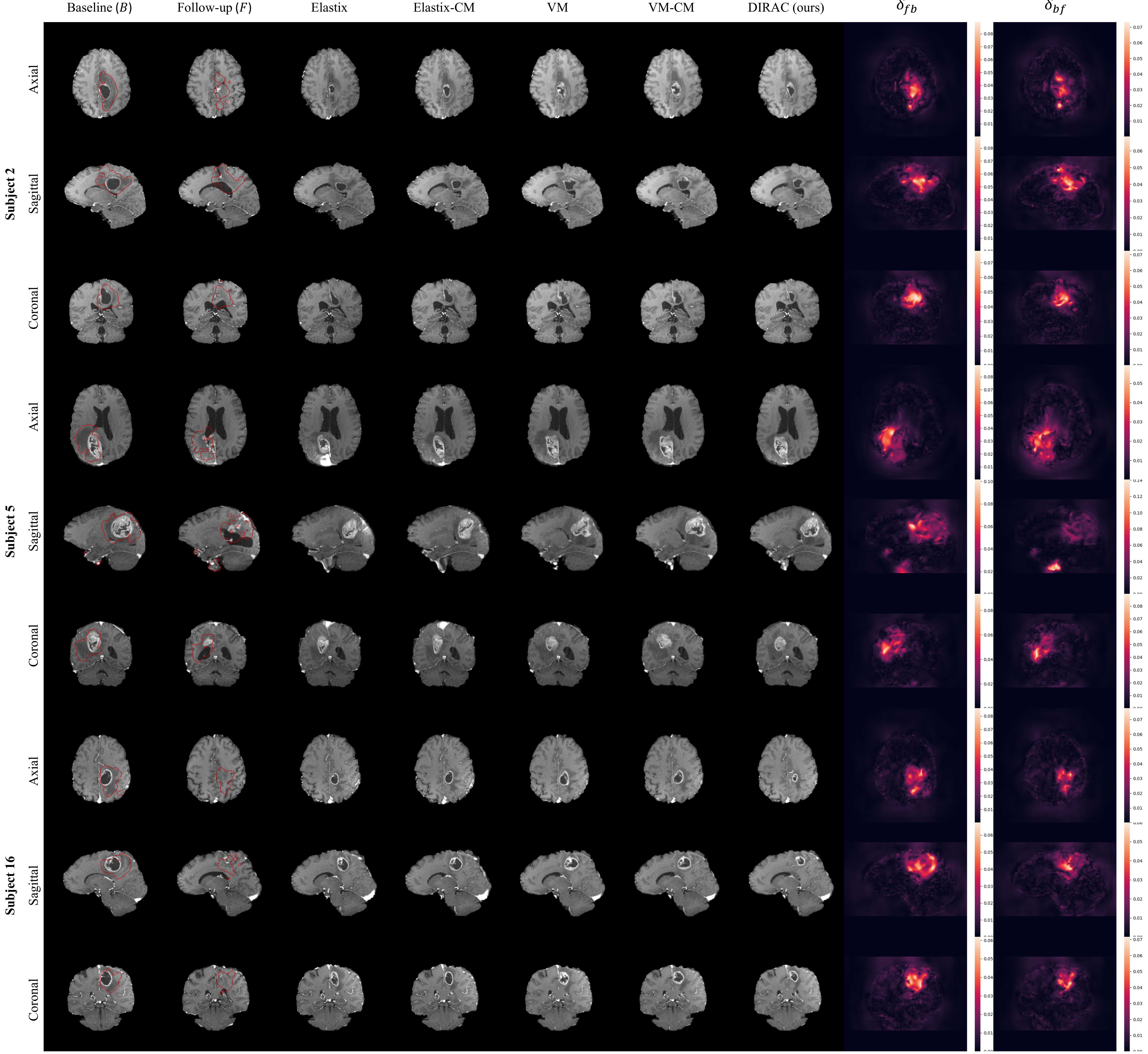}
	\end{center}
	\caption{Example axial, sagittal and coronal T1ce MR slices of resulting warped images ($B$ to $F$) from the baseline methods and our proposed method. The forward-backward errors ($\delta_{fb}$ and $\delta_{bf}$) of our method are shown next to our result. The estimated regions with absent correspondence from our method are overlaid with the baseline and follow-up scans (in red).}
	\label{fig:qualitative_a}
\end{figure}

\clearpage
\section{Distribution of target registration error}
\vspace{-15pt}
\begin{figure}[h]
	\centering
	\begin{tabular}{cc}
		
		\includegraphics[width=0.45\linewidth]{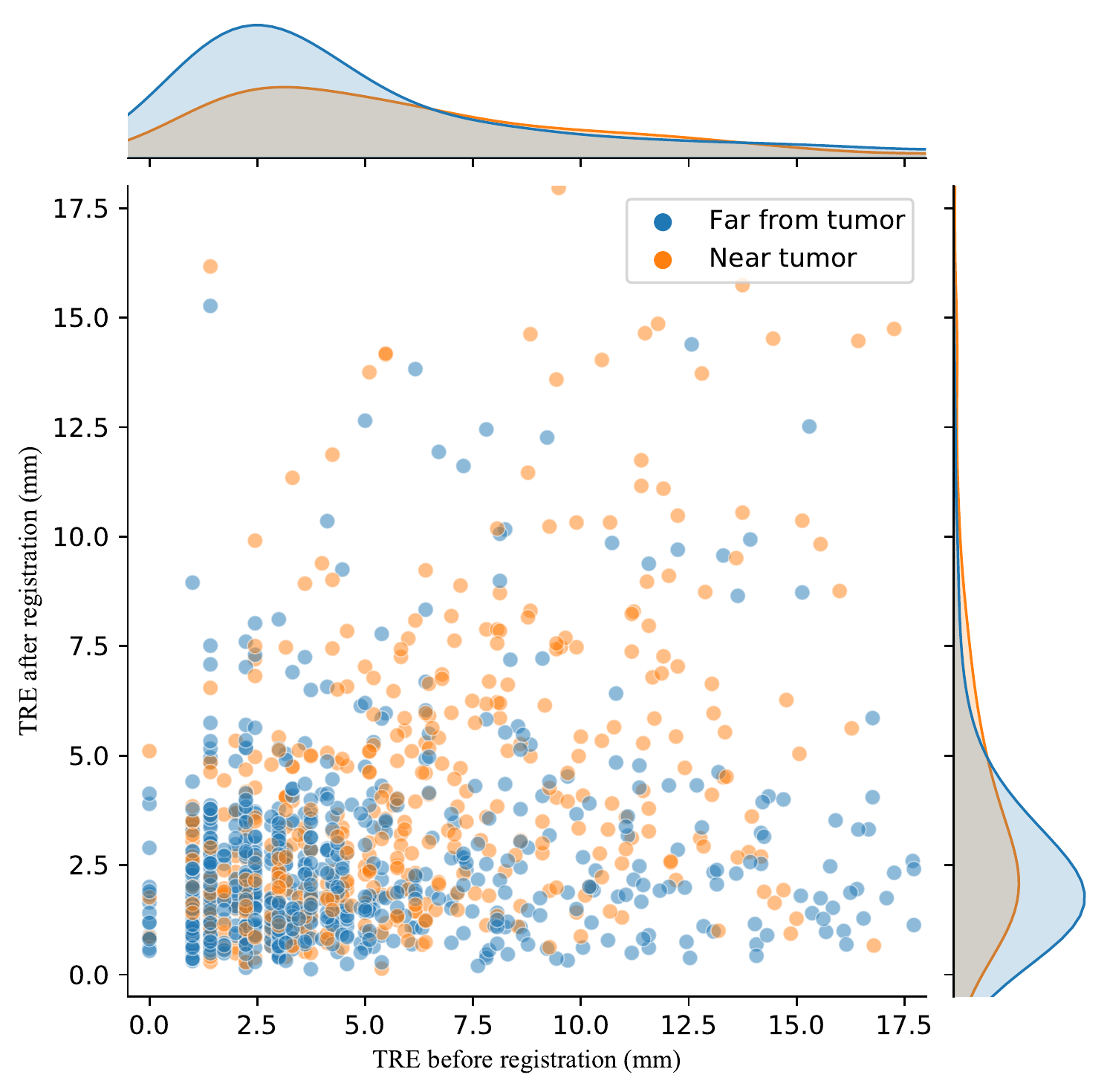} &
		\includegraphics[width=0.45\linewidth]{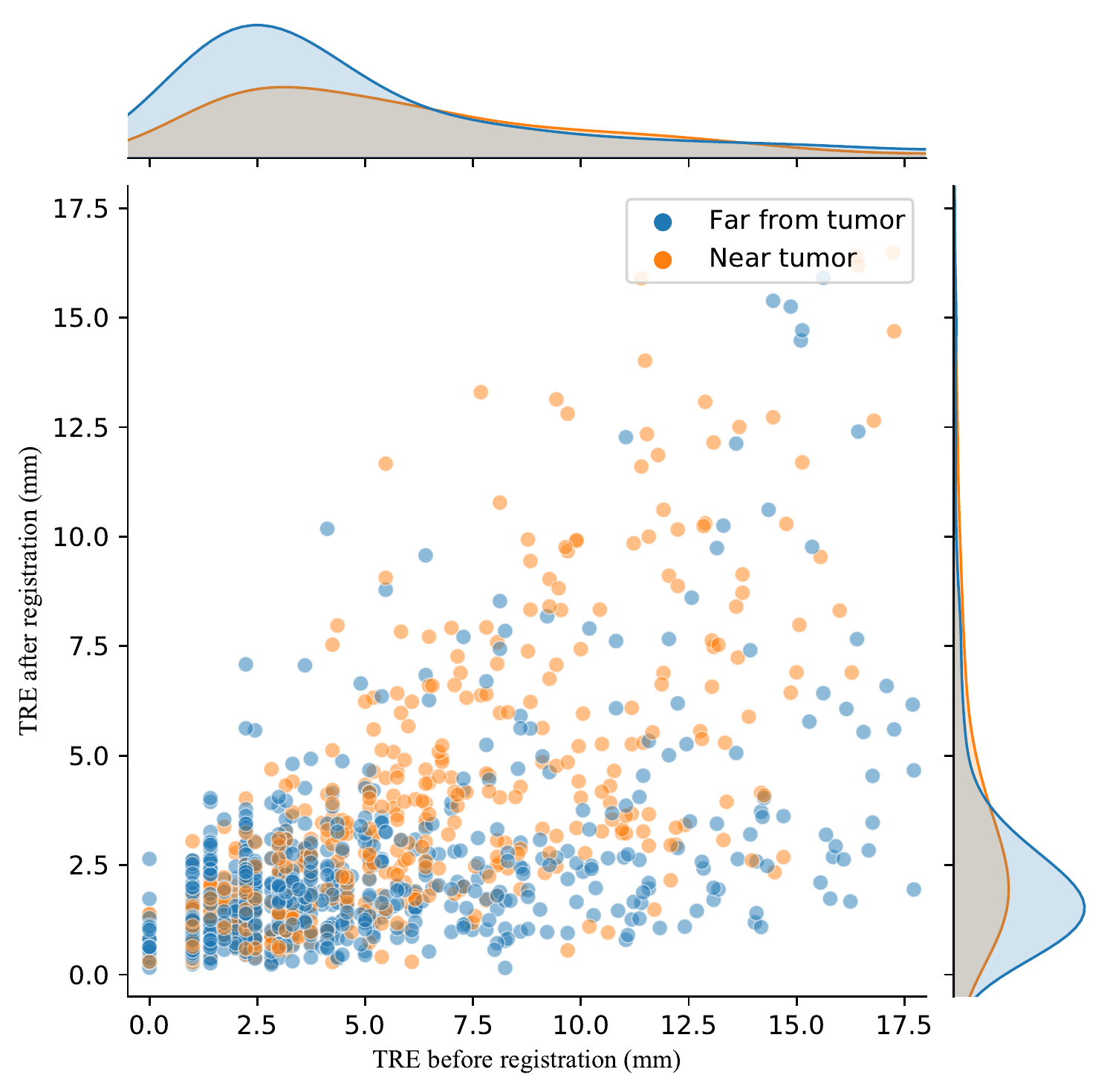} \\
		(a) Elastix  & (b) VM \\
		\includegraphics[width=0.45\linewidth]{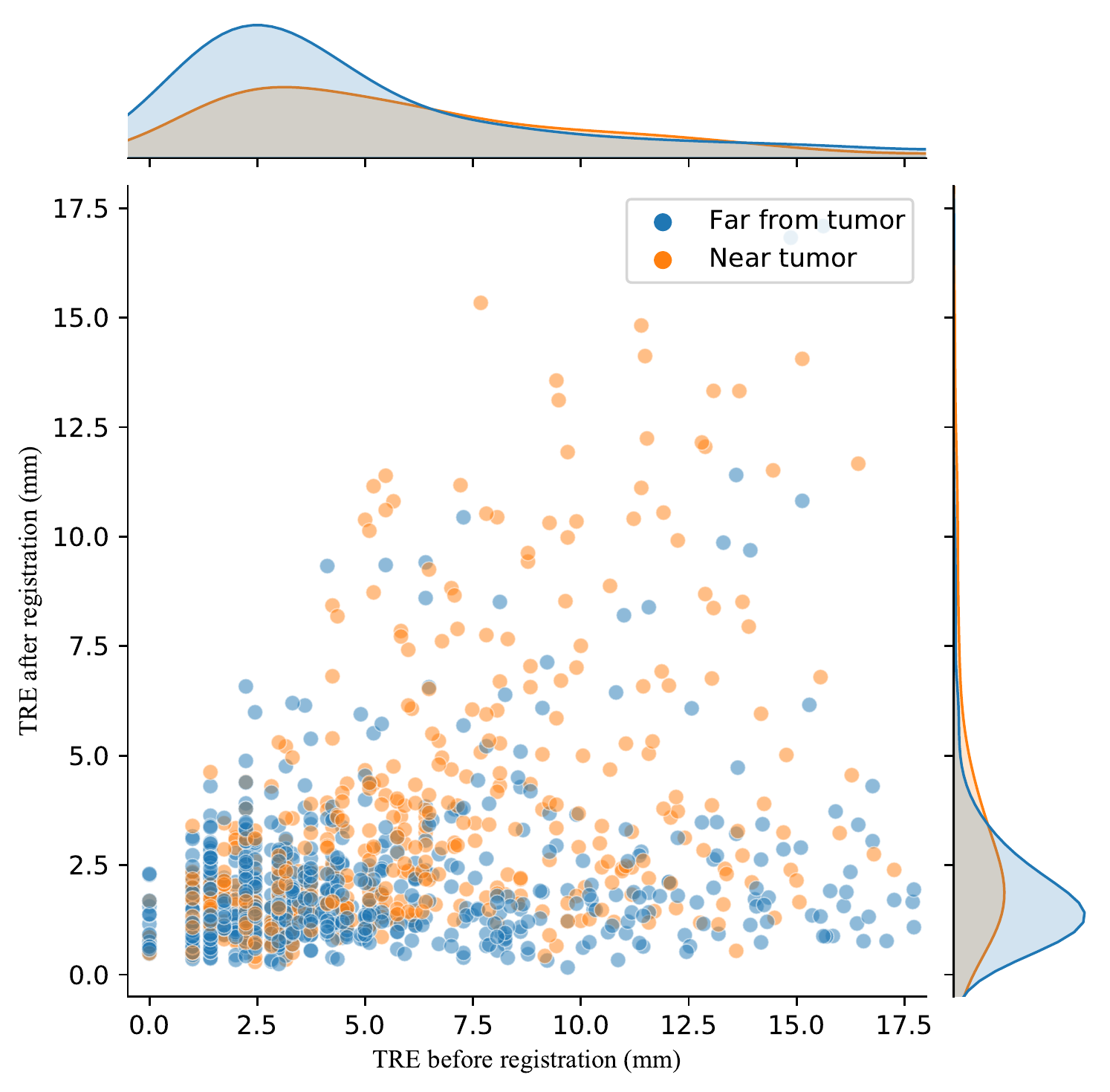} &
		\includegraphics[width=0.45\linewidth]{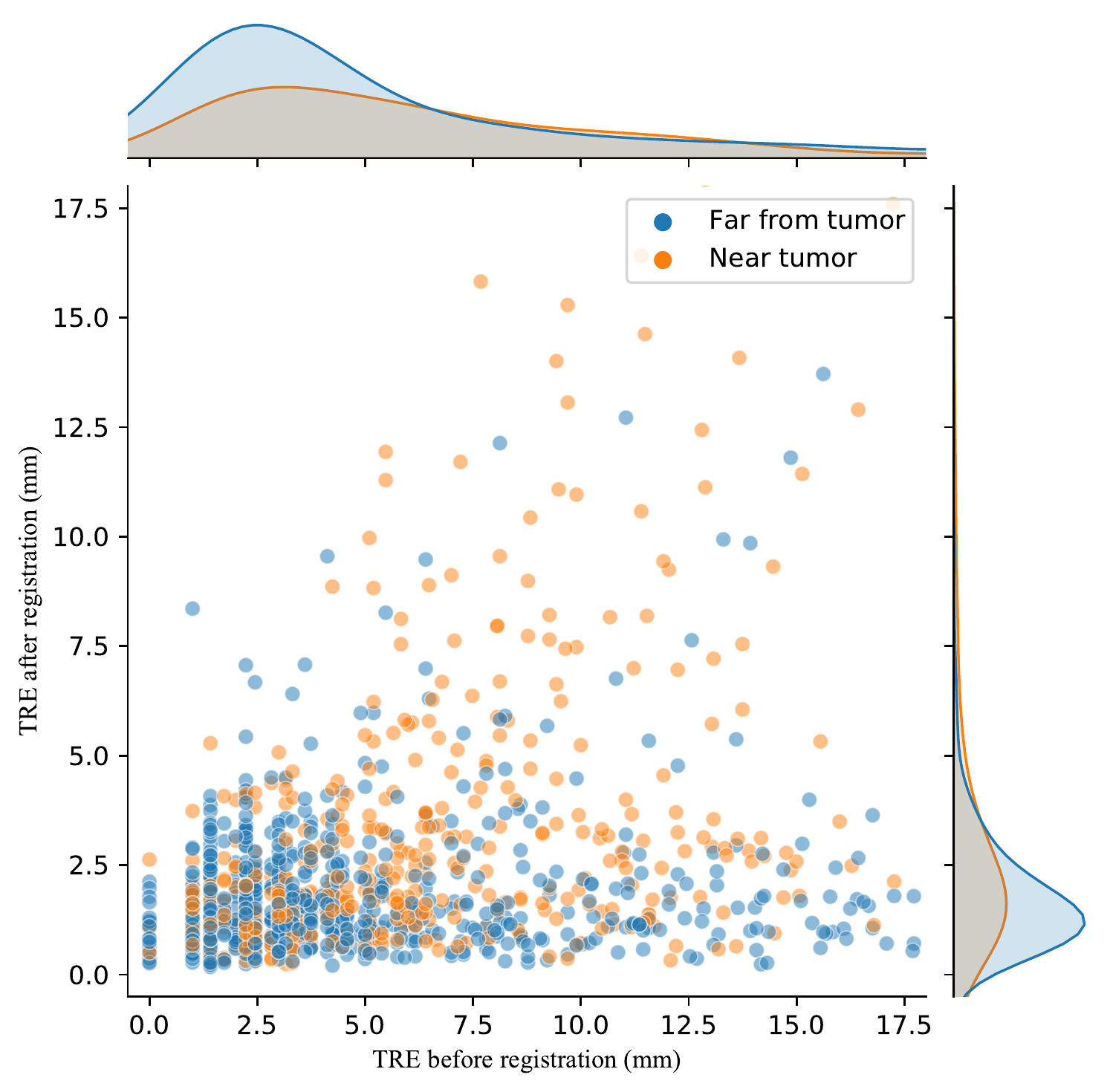} \\
		(c) cLapIRN  & (d) DIRAC (ours)
	\end{tabular}
	\caption{Scatterplots with joint histograms illustrate the target registration error (TRE) of the landmarks before and after registration using (a) Elastix, (b) VM, (c) cLapIRN and (d) our method. 80\% of landmarks after registration using our method is lower than 3.13 mm, while only 65\%, 70\% and 76\% of landmarks are lower than 3.13 mm after registration using Elastix, VM and cLapIRN, respectively.}
	
    \label{fig:quat_1}
\end{figure}

\clearpage

\end{appendix}

\end{document}